\documentclass[12pt]{article}

\catcode`@=12
\begin{document}
\thispagestyle{empty}
\begin{flushright} 
UCRHEP-T332\\ 
March 2002\
\end{flushright}
\vspace{0.5in}
\begin{center}
{\LARGE	\bf Uniqueness of a recently proposed\\ U(1) gauge extension\\}
\vspace{1.5in}
{\bf Ernest Ma\\}
\vspace{0.2in}
{Physics Department, University of California, Riverside, 
California 92521, USA\\}
\vspace{1.5in}
\end{center}
\begin{abstract}\
Consider the addition of a right-handed $SU(2)$ fermion multiplet (with 
neither color nor hypercharge) to each family of quarks and leptons.  The 
resultant theory admits a new U(1) gauge symmetry only if the additional 
multiplet is a singlet $N_R$, which corresponds to the well-known case of 
$U(1)_{B-L}$, or a triplet $(\Sigma^+, \Sigma^0, \Sigma^-)$, which 
corresponds to the proposal of hep-ph/0112232.  This disproves the 
assertion that the latter is somehow a ``trivial'' or ``expected'' discovery.
\end{abstract}
\newpage
\baselineskip 24pt

Consider $SU(3)_C \times SU(2)_L \times U(1)_Y \times U(1)_X$ as a possible 
extension of the standard model, under which each family of quarks and 
leptons transforms as follows:
\begin{eqnarray}
&& (u,d)_L \sim (3,2,1/6;n_1), ~~~ u_R \sim (3,1,2/3;n_2), ~~~ 
d_R \sim (3,1,-1/3;n_3), \nonumber \\ 
&& (\nu,e)_L \sim (1,2,-1/2;n_4), ~~~ e_R \sim (1,1,-1;n_5).
\end{eqnarray}
Add to these a right-handed fermion multiplet $(1,2p+1,0;n_6)$, where $p=0$ 
would correspond to a singlet, say $N_R$, as in the usual extension to 
include a right-handed neutrino singlet, and $p=1$ would correspond to a 
triplet $(\Sigma^+, \Sigma^0, \Sigma^-)_R$, as proposed recently \cite{new}.

As shown in Ref.~[1], there are 6 conditions to be satisfied for the gauging 
of $U(1)_X$.  Three of them do not involve $n_6$ and have 2 solutions:
\begin{eqnarray}
({\rm I}): && n_3 = 2n_1-n_2, ~~ n_4 = -3n_1, ~~ n_5 = -2n_1-n_2; \\ 
({\rm II}): && n_2 = {1 \over 4} (7n_1-3n_4), ~~ n_3 = {1 \over 4} (n_1+3n_4), ~~ 
n_5 = {1 \over 4} (-9n_1+5n_4).
\end{eqnarray}
The other 3 involve $n_6$, and they are given by
\begin{eqnarray}
{1 \over 2} (3n_1 + n_4) &=& {1 \over 3}p(p+1)(2p+1) n_6, \\ 
6n_1 - 3n_2 - 3n_3 + 2n_4 - n_5 &=& (2p+1) n_6, \\ 
6n_1^3 - 3n_2^3 - 3n_3^3 + 2n_4^3 - n_5^3 &=& (2p+1) n_6^3.
\end{eqnarray}

To find solutions to the above 3 equations, consider first $p=0$, then Eq.~(4) 
forces one to choose solution (I), and all other equations are satisfied 
with $n_1=n_2=n_3$ and $n_4=n_5=n_6$, i.e. $U(1)_{B-L}$ has been obtained. 
Consider now $p \neq 0$, then if solution (I) is again chosen, $n_6=0$ is 
required, which leads to $U(1)_Y$, so there is nothing new.

Now consider $p \neq 0$ and solution (II).  From Eqs.~(4), (5), and (6), it is 
easily shown that
\begin{equation}
{4n_6 \over 3n_1+n_4} = {6 \over p(p+1)(2p+1)} = {3 \over 2p+1} = 
\left( {3 \over 2p+1} \right)^{1\over 3},
\end{equation}
which clearly gives the unique solution of $p=1$, i.e. a triplet.  The fact 
that such a solution even exists (and for an integer value of $p$) for the 
above overconstrained set of conditions is certainly not a ``trivial'' or 
even ``expected'' result.

This work was supported in part by the U.~S.~Department of Energy under 
Grant No.~DE-FG03-94ER40837.

\bibliographystyle{unsrt}

\end{document}